\documentclass[submission, Phys]{SciPost}

\usepackage{mathtools}
\usepackage{bm,bbold,amsfonts,mathrsfs,amssymb,dsfont}

\usepackage{booktabs}
\usepackage{array}

\newcommand{\be}{\begin{equation}}
\newcommand{\ee}{\end{equation}}

\DeclareMathOperator{\tr}{\mbox{tr}}

\renewcommand{\L}{{\mathcal L}}

\newcommand{\nmax}{n_{\rm max}}

\newcommand{\p}{{\bf p}}

\newcommand{\ii}{{\mathbf i}}
\newcommand{\jj}{{\mathbf j}}

\begin{document}

\begin{center}{\Large \textbf{
Recursion method for quench dynamics:\\
strengths and limitations
}}\end{center}

\begin{center}
Ilya Shirokov\textsuperscript{1,2}, Viacheslav Khrushchev\textsuperscript{3}, Filipp Uskov\textsuperscript{1}, Ivan Dudinets\textsuperscript{8,4}, Igor Ermakov\textsuperscript{5,6,7,8}, Oleg Lychkovskiy\textsuperscript{1,7}
\end{center}

\begin{center}
{\bf 1} Skolkovo Institute of Science and Technology, \\ Bolshoy Boulevard 30, bld. 1, Moscow 121205, Russia
\\
{\bf 2} Moscow State University, Faculty of Physics,  Moscow 119991, Russia
\\
{\bf 3} HSE University,  20 Myasnitskaya Ulitsa, Moscow, 101000, Russia
\\
{\bf 4} Moscow Institute of Physics and Technology,\\ Institutskii per. 9, Dolgoprudnyi, 141700, Russia
\\
{\bf 5} Wilczek Quantum Center, Shanghai Institute for Advanced Studies, \\ University of Science and Technology of China, Shanghai 201315, China
\\
{\bf 6} Hefei National Laboratory,  Hefei 230088, China
\\
{\bf 7} Department of Mathematical Methods for Quantum Technologies, \\ Steklov Mathematical Institute of Russian Academy of Sciences, \\ 8 Gubkina St., Moscow 119991, Russia
\\
{\bf 8} Russian Quantum Center, Bolshoy Boulevard 30, bld. 1, Moscow 121205, Russia
\end{center}

\begin{center}
\today
\end{center}


\section*{Abstract}
{\bf
The recursion method, which solves coupled Heisenberg equations in a Lanczos operator basis, has recently emerged as a powerful nonperturbative tool for computing dynamical correlation functions in strongly correlated two- and three-dimensional quantum many-body systems. Motivated by this success, we investigate whether the method can be extended to expectation values of observables following a quantum quench. We find that such an extension encounters an obstacle absent in the computation of dynamical correlation functions. The latter are fully determined by the Lanczos coefficients~$b_n$, which in generic systems exhibit universal behavior, enabling reliable extrapolation from the first few dozens of explicitly computed coefficients. In contrast, quench dynamics additionally requires “quench coefficients”~$c_n$, defined as overlaps of Lanczos basis operators with the initial state. We show that, unlike the Lanczos coefficients, the quench coefficients exhibit no universal structure and cannot be reliably extrapolated, thereby limiting the time up to which the method yields accurate results. The behavior of quench coefficients is highly state-dependent, ranging from decaying to irregular or even growing sequences; typically, the less regular the sequence~$c_n$, the shorter the accessible timescale. Nevertheless, for favorable initial states, the method remains competitive with state-of-the-art approaches. Moreover, its symbolic implementation allows a single computation to be reused across different Hamiltonian parameters and initial states, making it particularly advantageous in studies requiring extensive scans over Hamiltonian parameters or initial states.
}

\section{Introduction}

Addressing quantum dynamics by means of the recursion method has a long history
\cite{Mori_1965_Continued-fraction,Dupuis_1967_Moment,Haydock_1980_Recursive,Mattis_1981_Reduce,Joslin_1986_Calculation,Roldan_1986_Dynamic,
viswanath2008recursion,Lindner_2010_Conductivity,Nandy_2025_Quantum}. The (Heisenberg version of the) method essentially amounts to constructing an orthonormal Lanczos basis within the Krylov space of an evolving quantum observable and harnessing the simplifications this basis brings about. A new impact has been given to the method by the universal operator growth hypothesis \cite{Parker_2019} -- a conjecture about a generic form of the evolution generator in the Lanczos basis.

So far, the practical application of the recursion method to quantum many-body systems has been limited almost exclusively to addressing  near-equilibrium dynamics -- specifically, computing  dynamical correlation functions and related transport properties. Several studies have reported such computations in the high-temperature limit \cite{Parker_2019,Khait_2016_Spin,deSouza_2020_Dynamics,Yates_2020_Lifetime,Yates_2020_Dynamics,Yuan_2021_Spin,Wang_2024_Diffusion,
Bartsch_2024_Estimation,Uskov_Lychkovskiy_2024_Quantum,Teretenkov_2025_Pseudomode,Bhattacharyya_2024_Metallic,Fullgraf_2025_Lanczos,ermakov2025symbolic}. A pathway to extend the approach to finite-temperature dynamical correlation functions has recently been charted \cite{Angelinos_2026_Temperature}. The efficiency of the method to compute infinite-temperature dynamical correlation functions in the nonperturbative regime is particularly notable in application to two-dimensional \cite{Uskov_Lychkovskiy_2024_Quantum,Teretenkov_2025_Pseudomode,Bhattacharyya_2024_Metallic} and three-dimensional \cite{ermakov2025symbolic} systems, where traditional numerical methods often struggle.  Another remarkable advantage of the  recursion method is that it allows a straightforward symbolic implementation, whereby a single computation covers all parameters of the Hamiltonian  \cite{Uskov_Lychkovskiy_2024_Quantum,Teretenkov_2025_Pseudomode}, providing a huge practical benefit over purely numerical methods.

Formally, the output of the recursion method is the Heisenberg operator of an observable. This suggests that the scope of the method can be much broader than near-equilibrium physics manifested by dynamical correlation functions and should cover  the  out-of-equilibrium dynamics following a quantum quench \cite{Loizeau_2025_Opening}. Addressing the latter in systems of dimension higher than one is known to be highly challenging, with only a limited body of results obtained to date \cite{Starkov_2018_Hybrid,Li_2020_Kosterlitz,Schmitt_2020_Quantum,Richter_2020_Quantum,Dziarmaga_2022_Time,Bernier_2024_Spatiotemporal,
Sinibaldi_2026_Time-Dependent,Gravina_2025_Neural,Pavesic_2025_Constrained,Chen_2026_Convolutional,
Begusic_2025_Real-time,Loizeau_2025_Quantum,Loizeau_2025_Codebase,rudolph2025pauli}.
Given the success of the recursion method in computing two- and three-dimensional dynamical correlation functions, it seems tempting to apply it to quench dynamics.

Here we undertake precisely this attempt. We perform case studies for one-, two- and three-dimensional spin-$1/2$ models. To this end, we leverage a novel software tool, \mbox{\textsc{QCommute}} \cite{lychkovskiy2026qcommute}, for the efficient symbolic computation of nested commutators.

We find that addressing post-quench out-of-equilibrium dynamics with the recursion method is substantially more difficult than computing dynamical correlation functions. We trace this difficulty to the nonuniversal, irregular, and, in general, unbounded behavior of the {\it quench coefficients} -- an ingredient absent in dynamical correlation functions but required for quench dynamics.

We show, however, that the recursion method can nevertheless be competitive with state-of-the-art approaches, both in accuracy and, to a lesser extent, in the maximum accessible timescale. Moreover, its symbolic implementation requires the core and most computationally demanding procedure to be performed only once. The resulting symbolic data remain valid across the full parameter space of the Hamiltonian and for arbitrary explicitly specified initial states, making the method particularly well suited for systematic scans over Hamiltonian parameters and initial states.

The paper is organized as follows. After setting the stage in the next section, we introduce the Heisenberg equation in the Lanczos basis -- the core of the recursion method -- in Sec.~\ref{sec: Lanczos basis}. In Sec.~\ref{sec: universal asymp}, we review the universal asymptotics of the Lanczos coefficients and explain how it enables the computation of dynamical correlation functions. We then apply the recursion method to quench dynamics in Sec.~\ref{sec: quench}. This is the central section of the paper: it introduces the quench coefficients, examines their nonuniversal behavior, demonstrates that this behavior fundamentally limits the applicability of the method to quench dynamics, and presents results for two- and three-dimensional spin-$1/2$ models. In Sec.~\ref{sec: comparison}, we compare the recursion-method results with those obtained using other state-of-the-art approaches. In Sec.~\ref{sec: 1D}, we separately consider the application of the method to a one-dimensional system. Finally, Sec.~\ref{sec: summary} summarizes our findings and outlines open questions and directions for future work.

\section{Setting the stage \label{sec: setting stage}}

Consider an observable given by a self-adjoint Schr\"{o}dinger operator~$A$. The same observable in the Heisenberg representation reads $A_t=e^{i t H} A \, e^{-i t H}$, where $H$ is the Hamiltonian of the system. The Heisenberg operator $A_t$ evolves within the {\it Krylov space} spanned by  operators $A,\L A,\L^2 A,\dots$, where
\begin{equation}
\L\equiv [H,\bullet]
\end{equation}
is the {\it superoperator} referred to as Liouvillian. $A_t$ satisfies the Heisenberg equation of motion
\begin{equation}\label{Heisenberg}
\partial_t A_t=i\, \L A_t,\quad A_{t=0}=A.
\end{equation}

If a system is initialized at $t=0$ in some nonequilibrium (pure or mixed) initial state $\rho_0$, the expectation value $\langle A_t \rangle$ of the observable $A$ evolves according to
\begin{equation}\label{quench}
\langle A_t \rangle=\tr \big(\rho_0 A_t\big).
\end{equation}
This setting is traditionally referred to as quantum quench.

One can Taylor-expand $\langle A_t \rangle$ to obtain
\begin{equation}\label{Taylor}
\langle A_t \rangle=\sum_{n=0}^\infty \frac{(it)^n}{n!}\tr \big( \rho_0\, \L^n A\Big).
\end{equation}
For many-body systems in the thermodynamic limit, this expansion typically has a finite convergence radius, except in one dimension \cite{Parker_2019}.

To illustrate our key findings, we focus on a specific  spin-$1/2$ model in two or three dimensions -- the quantum Ising model on the square or cubic lattice, respectively.  Its Hamiltonian reads
\begin{equation}\label{H Ising}
H=-\sum_{\langle \ii \jj \rangle} \sigma^x_{\ii}\sigma^x_{\jj}-h_z \sum_{\jj} \sigma^z_{\jj},
\end{equation}
where  $\sigma^{x,y,z}_{\jj}$ are Pauli matrices, $\ii$ and $\jj$ enumerate lattice sites and the first sum runs over pairs of neighbouring sites. We will also consider a one-dimensional tilted-field Ising model, which will be separately discussed in Sec. \ref{sec: 1D}.

\begin{figure}[th!] 
		\centering
    	\includegraphics[width=\textwidth]{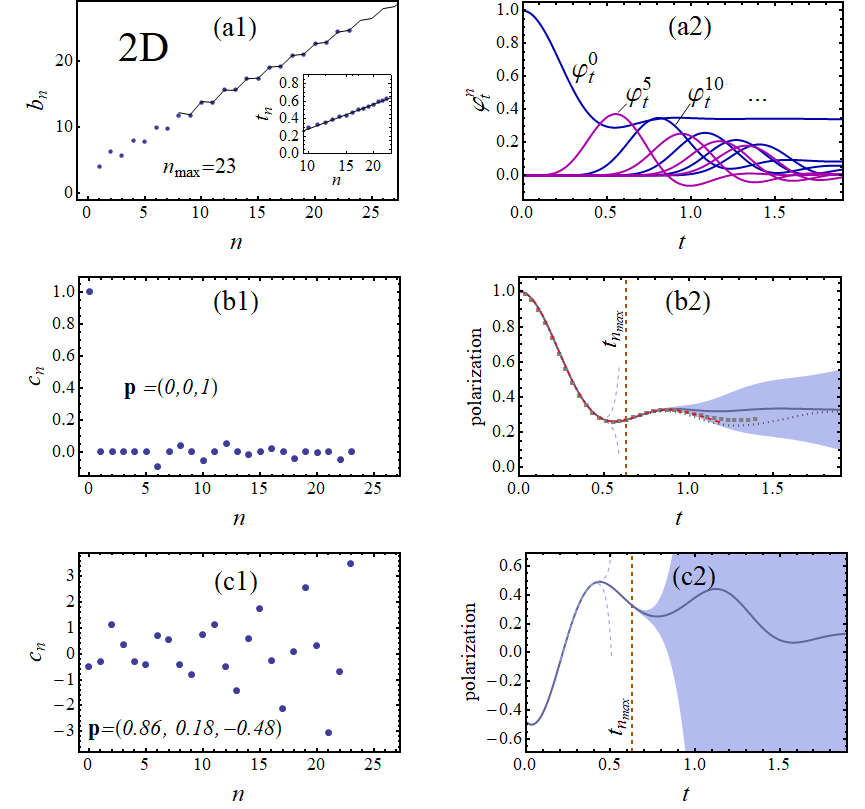}
		\caption{Quench dynamics of the per-spin polarization in the $z$-direction for the Ising model \eqref{H Ising} on the square lattice  with $h_z=1$. {\bf (a1).} Explicitly computed Lanczos coefficients (points) and their extrapolation (line). Inset shows the scaling of time $t_n$ with $n$ (see text for details). {\bf (a2).} Functions $\varphi^n_t$ entering the expansion \eqref{At} of the Heisenberg operator. Blue (magenta) colour indicates even (odd) $n$.  {\bf (b1),(c1).} Quench coefficients for the product initial state  \eqref{initial state} with the initial polarizations $\p$ indicated in the plots. The two plots demonstrate two different qualitative behaviors: decaying (b1) and unbounded (c1) quench coefficients.  {\bf (b2),(c2).} Blue solid line -- dynamics of the per-spin polarization in the $z$-direction approximated by the truncated sum \eqref{At quench truncated}. Shaded area indicates the uncertainty due to the truncation estimated according to eq. \eqref{truncation uncertainty}. Blue dashed lines show Taylor series~\eqref{Taylor} truncated at orders $\nmax$ and $(\nmax-2)$. Gray squares and red dashed line in (b2) show the results of sparse Pauli dynamics \cite{Begusic_2025_Real-time,Begusic_2025_unpublished} and numerical linked cluster expansion \cite{Richter_2020_Quantum}, respectively. Additionally, the numerically exact result for a finite $5\times5$ system with periodic boundary conditions~\cite{Richter_2020_Quantum} is shown by the gray dotted line.}
		\label{fig 2D}
\end{figure}

\begin{figure}[t] 
		\centering
    	\includegraphics[width=\linewidth]{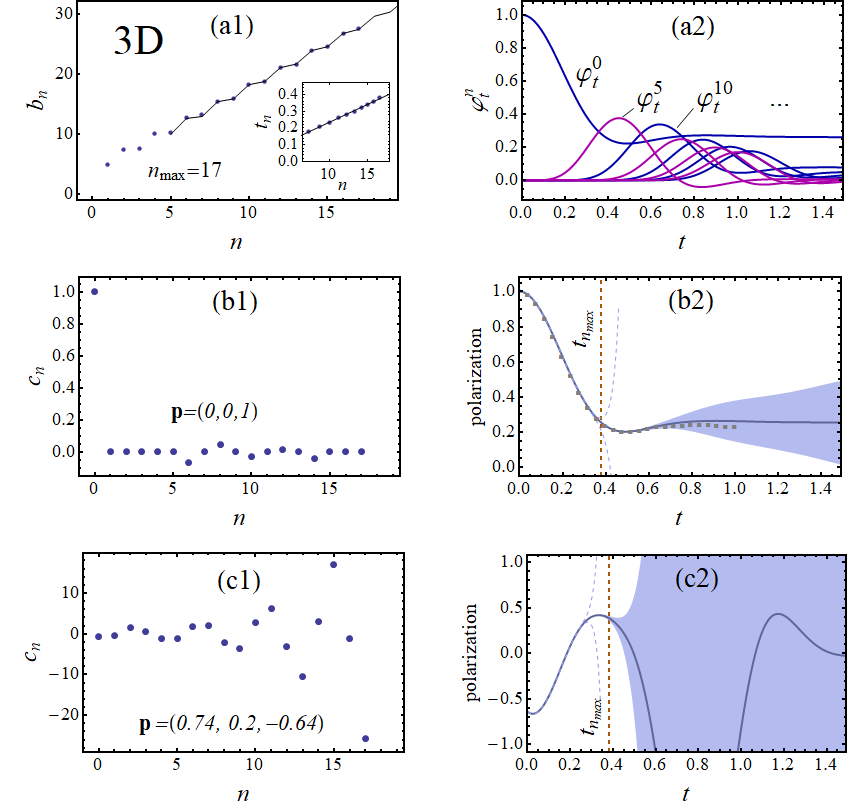}
		\caption{The same as in Fig. \ref{fig 2D}, but for the Ising model \eqref{H Ising} on the cubic lattice with $h_z=1$. }
		\label{fig 3D}
\end{figure}

The observable we consider is the total polarization in the $z$-direction,
\begin{equation}\label{observable}
A=\sum_{\jj} \sigma^z_{\jj}.
\end{equation}
Since this observable is extensive in the system size (number of spins), we will present the results in Figs.~\ref{fig 2D}--\ref{fig 1D} in terms of the per-spin polarization $N^{-1}\langle A_t \rangle$.

We focus on  product translation-invariant initial states with polarization $\p$,
\begin{equation}\label{initial state}
\rho_0=\bigotimes_j \frac12(\mathbb{1}+\p \boldsymbol{\sigma}_j),
\end{equation}
where $\p \boldsymbol{\sigma}_j\equiv p_x\sigma^x_j+p_y\sigma^y_j+p_z\sigma^z_j$. In general, $|\p|\leq 1$; in particular, $|\p|= 1$ for pure states.

\section{Heisenberg equation in a Lanczos basis \label{sec: Lanczos basis}}

Within the recursion method  \cite{viswanath2008recursion}, the Heisenberg equation \eqref{Heisenberg} is addressed in the orthonormal Lanczos basis of operators, $\{Q_0,Q_1,Q_2,\dots\}$. To define this basis, one first needs to choose a scalar product between operators. We adopt the most straightforward scalar product,
\begin{equation}\label{scalar product}
\big(A|B\big)\equiv {\cal N}\,\tr \big(A^\dagger B\big),
\end{equation}
where ${\cal N}$ is a normalization constant. The scalar product entails the norm $\|A\|=\sqrt{(A|A)}$.
For systems of $N$ spins $1/2$ and translation-invariant observables $A$ considered here it is natural to choose ${\cal N}=N^{-1}\,2^{-N}$, in which case $\|A\|$ is independent of the system size. The advantage of the scalar product \eqref{scalar product} is that it can be readily computed for any explicitly given operators.

The orthonormal Lanczos basis $\{Q_0,Q_1,Q_2,\dots\}$ is constructed iteratively as follows:
\begin{align}
&&&&Q_0 &= A / \|A\|, & \\
A_1 &= i\L Q_0,                         &b_1 &= \|A_1\|, & Q_1 &= b_1^{-1} A_1,  \nonumber\\
A_2 &= i\L Q_1 + b_1 Q_0,   &b_2 &= \|A_2\|, & Q_2 &= b_2^{-1} A_2, \nonumber\\
&\qquad\qquad\qquad\qquad\quad\dots \nonumber \\
A_n &= i\L Q_{n-1} + b_{n-1} Q_{n-2},   &b_n &= \|A_n\|, & Q_n &= b_n^{-1} A_n, \nonumber\\
&\qquad\qquad\qquad\qquad\quad\dots \nonumber
\end{align}
Here $A_n$ are unnormalized counterparts of $Q_n$, and coefficients $b_n$ are known as {\it Lanczos coefficients}. Each $Q_n$ is self-adjoint by construction.\footnote{Often one uses $O_n\equiv (-i)^nQ_n$ as basis operators. $O_n$ are not self-adjoint for odd $n$ -- for this reason we find using $Q_n$ more convenient. } Note that each observable $A$ defines its own Lanczos basis.


In the Lanczos basis, the Liouvillian admits a tridiagonal form: $(Q_{n+1}|i\L|Q_n)=b_{n+1}$, $(Q_{n-1}|i\L|Q_n)=-b_n$ and other matrix elements vanish. In a slight abuse of notations, we use the same symbol $\cal L$ for the matrix form of the Liouvillian in the Lanczos basis:
\begin{equation}\label{L}
i\L=
\begin{pmatrix}
   0  & -b_1 & 0  & 0  & \dots  \\[0.3 em]
   b_1 & 0  & - b_2 & 0  & \dots \\[0.3 em]
     0 & b_2  & 0 &  - b_3  &  \dots \\[0.3 em]
     0 & 0  & b_3 & 0  &  \dots \\[0.3 em]
\vdots & \vdots & \vdots & \vdots & \ddots
  \end{pmatrix}.
\end{equation}

The Heisenberg equation \eqref{Heisenberg} can be rewritten in the matrix form as
\begin{align}\label{Heisenberg equation matrix form}
    \partial_t \varphi_t  &=i\L\varphi_t,
\end{align}
where $i\cal L$ is the tridiagonal matrix \eqref{L} and $\varphi_t=\|\varphi^n_t\|_{n=0,1,2,\dots}$ is a column vector consisting of functions  \begin{equation}
\varphi^n_t\equiv (Q_n|A_t)/\|A\|.
\end{equation}
The initial condition for the equation \eqref{Heisenberg equation matrix form} reads
\begin{equation}
\varphi^n_{t=0}=\delta_{n0}.
\end{equation}

Note that  $\varphi^0_t =\tr A\,A_t/\tr A^2$ is the normalized dynamical autocorrelation function of the observable $A$ in the limit of the infinite temperature.

Once the Lanczos basis is constructed and eq.\eqref{Heisenberg equation matrix form} is solved, the solution of the Heisenberg equation in its original form~\eqref{Heisenberg} is given by
\begin{equation}\label{At}
A_t=\|A\|\,\sum_{n=0}^{\infty} \varphi^n_t \,  Q_n.
\end{equation}


\section{Extrapolating Lanczos coefficients and computing $\varphi^n_t$ \label{sec: universal asymp}}

In practice, in the many-body case one can explicitly compute only a limited number $\nmax$ of nested commutators ${\cal L}^n A$ and, hence, Lanczos coefficients $b_n$, so most of the matrix~\eqref{L} remains unknown. Fortunately, often the Lanczos sequence $b_n$ can be reasonably extrapolated beyond $\nmax$. That such extrapolation is generally possible has been first demonstrated in ref.~\cite{Parker_2019}, in which the universal operator hypothesis (UOGH) has been put forward (see also precursor work \cite{Liu_1990_Infinite-temperature,Florencio_1992_Quantum,Zobov_2006_Second,Elsayed_2014_Signatures,Bouch_2015_Complex}). The hypothesis states  that, for a generic many-body system in the thermodynamic limit, the Lanczos coefficients $b_n$ scale asymptotically linearly with $n$ at large $n$ (with a logarithmic correction for one-dimensional systems). This asymptotic behavior has been confirmed in various many-body models \cite{Cao_2021_Statistical,Noh_2021,Heveling_2022_Numerically,Uskov_Lychkovskiy_2024_Quantum,De_2024_Stochastic,ermakov2025operator,ermakov2025symbolic}. As a rule, the leading asymptotics can be easily extracted from the first dozen of Lanczos coefficients. However, it turns out that extrapolating  Lanczos coefficients by a linear function is a too rough approximation: the qualitative behavior of the solutions of the Heisenberg equation strongly depends on the subleading asymptotic terms \cite{Viswanath_1994_Ordering,Parker_2019,Yates_2020_Lifetime,Yates_2020_Dynamics,
Dymarsky_2021_Krylov,Bhattacharjee_2022_Krylov,Avdoshkin_2024_Krylov,Camargo_2023_Krylov,Uskov_Lychkovskiy_2024_Quantum,dodelson2025black,Gamayun_2025_Exactly}. Guided by the results of ref. \cite{Gamayun_2025_Exactly} (see also \cite{Lunt_2026_Emergent}), we employ the following extrapolation formula:
\begin{align}\label{b_n extrapolation 2D}
  b_n & \simeq  \alpha\, n + \gamma + \frac{(-1)^n \tilde\alpha }{\log n+c}.
\end{align}
The last alternating term is required whenever $\varphi^{0}_t$ relaxes to a nonzero value at $t\rightarrow \infty$ \cite{Gamayun_2025_Exactly}, which is generally the case.

Note that in a system with a finite dimension of the Hilbert space the linear growth of $b_n$ extends only up to a threshold $n$ which is linear in the system size. Beyond this threshold, $b_n$ approximately saturates, which is a finite-size effect \cite{Barbon_2019_On_evolution,Rabinovici_2021_Operator,Noh_2021}. Throughout the paper we work in the thermodynamic limit, i.e. we ignore finite-size effects and assume that eq. \eqref{b_n extrapolation 2D} is a true asymptotic of an infinite sequence of $b_n$.

The assumption about the asymptotic behavior of the Lanczos sequence considerably strengthens the capability of the recursion method  \cite{Parker_2019}. In our implementation of the method we proceed as follows \cite{Uskov_Lychkovskiy_2024_Quantum,ermakov2025symbolic}. The first step is to compute $\nmax$ nested commutator ${\cal L}^n A$. To this end we employ a newly developed, performance-optimized software tool \textsc{QCommute} \cite{lychkovskiy2026qcommute} that computes nested commutators for spin-$1/2$ many-body models on hypercubic lattices. The computation is performed directly in the thermodynamic limit and in the symbolic form (i.e. model parameters entering the Hamiltonian are kept symbolic). From the computational perspective, this is the most resource-consuming step. Next, $\nmax$ basis operators $O^n$ and Lanczos coefficients $b_n$ are computed for a specific set of model parameters. This is done with the use of rational arithmetics, which completely avoids the well-known numerical instability of the Lanczos algorithm \cite{Eckseler_2025_Escaping}.  Then the Lanczos sequence $b_n$ is extrapolated according to extrapolation formula \eqref{b_n extrapolation 2D}, with the parameters $\alpha,\,\tilde\alpha,\,\gamma$ and $c$  found by fitting $n_{\rm max}$ explicitly computed Lanczos coefficients. Finally, coupled differential equations \eqref{Heisenberg equation matrix form} are truncated at some large $n$ and solved numerically. The truncation order is chosen large enough (typically, on the order of $10^3$) that the truncation does not affect $\varphi^n_t$ up to the thermalization time. This implementation of the recursion method has proven highly effective in computing dynamical autocorrelation functions in two- and three-dimensional systems of spins $1/2$ and fermions \cite{Uskov_Lychkovskiy_2024_Quantum,Teretenkov_2025_Pseudomode,ermakov2025symbolic}. The resulting Lanczos coefficients and functions $\varphi^n_t$ for the two- and three-dimensional Ising model are shown in the upper rows of Figs.~\ref{fig 2D} and \ref{fig 3D}, respectively.

\section{Quantum quench via recursion method \label{sec: quench}}

In the framework of the recursion method, quench dynamics of the observable $A$ follows  from eq. \eqref{At}:
\begin{equation}\label{At quench}
\langle A_t \rangle=\|A\|\,\sum_{n=0}^{\infty} c_n\, \varphi^n_t,
\end{equation}
where the real numbers
\begin{equation}\label{quench coefficients}
c_n \equiv \tr \rho_0 Q_n
\end{equation}
are referred to as  {\it quench coefficients}. Quench coefficients can be viewed as the expansion coefficients of the initial density operator $\rho_0$ over the Lanczos basis (up to an overall normalization).

Quench coefficients $c_n$ arise as an additional ingredient specific to quench dynamics and absent in the computation of autocorrelation functions. Within the recursion method, one can explicitly compute the same number $\nmax$ of quench coefficients as Lanczos coefficients $b_n$. However, we find that, unlike the Lanczos sequence $b_n$, the sequence $c_n$ exhibits no apparent structure and no universal asymptotic behavior, and therefore cannot, in general, be meaningfully extrapolated beyond $\nmax$. Exemplary quench coefficients for the two- and three-dimensional Ising model for various initial states are shown in Figs.~\ref{fig 2D}(b1,c1) and \ref{fig 3D}(b1,c1), respectively.

Furthermore, we observe that the behavior of $c_n$ strongly depends on the initial state. For some initial states, the coefficients display a clear decay with $n$, whereas for others they exhibit irregular behavior and even grow with $n$. These two qualitatively distinct regimes are illustrated in Figs.~\ref{fig 2D}(b1), \ref{fig 3D}(b1) on the one hand, and Figs.~\ref{fig 2D}(c1), \ref{fig 3D}(c1) on the other. Taken together, these observations about the quench coefficients are central to this work.

Given the irregular nature of $c_n$, we now analyze the convergence of the sum in eq.~\eqref{At quench} and the error introduced by truncating it at $n=\nmax$, which is the only practical way to evaluate eq.~\eqref{At quench}:
\begin{equation}\label{At quench truncated}
\langle A_t \rangle \simeq \|A\|\,\sum_{n=0}^{\nmax} c_n\, \varphi^n_t.
\end{equation}

We first note that functions $\varphi^n_t$ are exponentially suppressed up to a time scale $t_n \sim \log n$, see Refs.~\cite{Parker_2019,Gamayun_2025_Exactly} and Figs.~\ref{fig 2D}(a2), \ref{fig 3D}(a2). As a consequence, the truncated expression \eqref{At quench truncated} is highly accurate for $t \leq t_{\nmax}$.

More precisely, we define $t_n$ as the time at which $|\varphi^n_t|$ first reaches a small threshold $\epsilon=10^{-3}$. One expects $t_n$ to increase with $n$  as $\log n$  \cite{Parker_2019,Gamayun_2025_Exactly}, and this scaling is confirmed by our numerical data (see insets in Figs.~\ref{fig 2D}(a1), \ref{fig 3D}(a1)).

The growth of $t_{\nmax}$ with $\nmax$, although only logarithmic, should be contrasted with the finite convergence radius of the truncated Taylor expansion \eqref{Taylor}. As shown in Figs.~\ref{fig 2D},\ref{fig 3D}, for currently accessible $\nmax$ the recursion method already reaches substantially longer times than the truncated Taylor expansion.

To assess the accuracy at later times, we introduce a quantitative estimate of the truncation error. We define the root-mean-square value $\overline c$ of the upper half of available coefficients, $\overline c^2=(1/\lceil\nmax/2\rceil)\sum\limits_{n=\lceil\nmax/2\rceil}^{\nmax} c_n^2$, where $\lceil x\rceil$ denotes the ceiling function, and estimate upper and lower bounds for $\langle A_t \rangle$ as
\begin{equation}\label{truncation uncertainty}
\langle A_t \rangle^{\pm}=\|A\|\left(\sum_{n=0}^{\nmax} c_n\, \varphi^n_t \pm \, \sum_{n=\nmax+1}^{\infty}\overline c\,\varphi^n_t\right).
\end{equation}
In practice, due to the suppression of $\varphi^n_t$ for $t<t_n$, only a finite number of terms contribute significantly to the second sum at any finite time. The resulting bounds are shown in Figs.~\ref{fig 2D}, \ref{fig 3D} as shaded regions indicating the estimated uncertainty.

With this estimate, we can examine the accuracy of the truncated expression \eqref{At quench truncated} beyond $t_{\nmax}$. We find that the accuracy strongly depends on the behavior of the quench coefficients. If $c_n$ are small around $n\simeq \nmax$ and continue to decrease, the truncation can remain accurate well beyond $t_{\nmax}$, as illustrated in Figs.~\ref{fig 2D}(b2), \ref{fig 3D}(b2). In contrast, if $c_n$ grow with $n$, the truncated expression becomes unreliable immediately beyond $t_{\nmax}$, as seen in Figs.~\ref{fig 2D}(c2), \ref{fig 3D}(c2). The example in Fig.~\ref{fig 3D}(c2) demonstrates a particularly severe breakdown, where the approximate per-spin polarization drops below $(-1)$, violating the physical constraint that its absolute value cannot exceed one.

In Appendix~\ref{appendix: zero energy}, we present results for the same models but for different initial states whose energy density corresponds to the infinite-temperature sector of the spectrum. These results, shown in Figs.~\ref{fig 2D E0} and \ref{fig 3D E0}, are qualitatively similar to those discussed in the present section. This rules out a possibility that the emergence of divergent sequences of $c_n$ and the associated limitations of the recursion method are due to a mismatch between the effective temperature of the initial state and the infinite temperature implicitly associated \cite{Parker_2019} with the choice \eqref{scalar product} of the scalar product.

We conclude this section with a general argument indicating that quench coefficients cannot decay too rapidly with $n$. Consider the sum rule
\begin{align}\label{sum rule failed}
\sum_{n=0}^\infty c_n^2 & = \sum_{n=0}^\infty \left(N\,2^N\right)^2\, (\rho_0|O^n) (O^n|\rho_0) \nonumber\\
                        & = \left(N\,2^N\right)^2\, (\rho_0^K | \rho_0^K )= N\,2^N \tr (\rho_0^K)^2,
\end{align}
where $\rho_0^K$ is the projection of $\rho_0$ on the Krylov subspace. Noting that $\tr \rho_0^2>2^{-N}$ and assuming that a similar bound (up to a factor independent of $N$) holds for $\rho_0^K$, one finds that the above sum diverges in the thermodynamic limit of $N\rightarrow\infty$. This implies that the coefficients $c_n$ cannot decay faster than $1/\sqrt{n^{1+\varepsilon}}$, where $\varepsilon>0$ is arbitrarily small.

\section{Comparison to other methods \label{sec: comparison}}

\begin{figure}[t!] 
		\centering
    	\includegraphics[width=\linewidth]{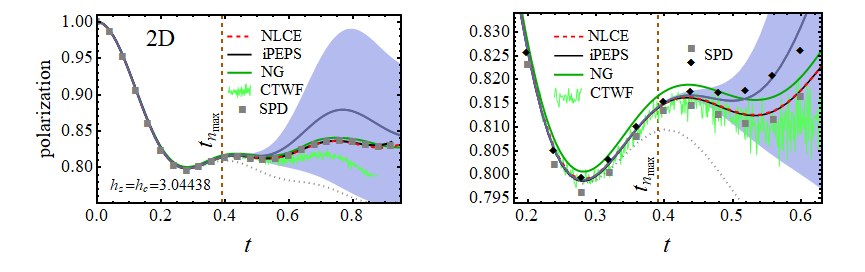}
		\caption{Comparison of the recursion method to other methods. Shown are results for the dynamics of the per-spin polarization in the $z$-direction for the Ising model \eqref{H Ising} on the square lattice at the critical field $h_z=3.04438$. The initial state is completely polarized in the $z$-direction, $\p=(0,0,1)$.   The right panel is a zoom of the left panel into the region where the comparison is most instructive.  Analogously to Figs. \ref{fig 2D},\ref{fig 3D}, blue solid line denotes the recursion method result, with shaded area indicating the estimated uncertainty.  For comparison, shown are results obtained using other methods: red dashed line -- NLCE~\cite{Richter_2020_Quantum}, black line -- iPEPS~\cite{Dziarmaga_2022_Time}, green line -- NG~\cite{Sinibaldi_2026_Time-Dependent}, green points -- CTWF~\cite{Chen_2026_Convolutional}, black diamonds -- SPD with Trotter step $\delta t=0.001$ and accuracy parameter $\delta=10^{-19}$, and gray squares -- SPD with $\delta t=0.004$ and $\delta=10^{-23}$~\cite{Begusic_2025_Real-time}. Note that NLCE and iPEPS results are indistinguishable up to $t\simeq 0.8$. The gray dotted line shows the numerically exact result for a finite $5\times 5$ lattice with periodic boundary conditions \cite{Richter_2020_Quantum}.}
		\label{fig comp}
\end{figure}

In this section, we compare the results obtained with the recursion method to those of other approaches. We begin with the two-dimensional case. We consider a setup that is widely studied and commonly used as a benchmark to test new methods: the quench from a fully polarized state in the Ising model \eqref{H Ising} on the square lattice at the critical field $h_z=3.04438$. We compare our results with those obtained using numerical linked cluster expansion (NLCE)~\cite{Richter_2020_Quantum}, infinite projected entangled pair states (iPEPS)~\cite{Dziarmaga_2022_Time}, the time-dependent neural Galerkin (NG) method~\cite{Sinibaldi_2026_Time-Dependent},  the time-dependent variational principle with convolutional transformer wave functions (CTWF)~\cite{Chen_2026_Convolutional}, and the sparse Pauli dynamics (SPD)~\cite{Begusic_2025_Real-time}; see Fig.~\ref{fig comp}. We briefly summarize these methods before discussing the comparison.

The NLCE method expands the expectation value of an observable in contributions from finite connected clusters, evaluating each contribution by Chebyshev real-time propagation on the corresponding cluster~\cite{Richter_2020_Quantum}. Ref.~\cite{Richter_2020_Quantum} also reports the numerically exact result for a finite $5\times5$ system with periodic boundary conditions, obtained using the same propagation scheme.

The iPEPS approach~\cite{Dziarmaga_2022_Time} performs a Trotterized real-time evolution of a tensor-network approximation to the many-body state directly in the thermodynamic limit.

The NG method~\cite{Sinibaldi_2026_Time-Dependent} implements a time-nonlocal variational principle within a manifold of neural quantum states, with results reported for the $8\times 8$ square lattice.

The CTWF approach~\cite{Chen_2026_Convolutional} employs the time-dependent variational principle with convolutional transformer wave functions, also on the $8\times 8$ square lattice.

SPD~\cite{Begusic_2025_Real-time} is, like the recursion method, formulated in the Heisenberg picture. It performs a Trotterized evolution of the observable on an $11\times 11$ lattice. To control operator growth, Pauli strings with amplitudes below a threshold $\delta$ are discarded. Both the finite Trotter time step $\Delta t$ and the truncation threshold $\delta$ introduce errors; for a fixed computational budget, improving one source of error typically requires worsening the other, resulting in a trade-off between short-time and long-time accuracy \cite{Park_2025_Simulating}.

Among the methods considered, only recursion method, NLCE and iPEPS operate directly in the thermodynamic limit, whereas the others are restricted to finite systems. However, it has been shown that for lattice sizes $8\times 8$ and larger, finite-size effects set in only at times $t \gtrsim 0.6$~\cite{Park_2025_Simulating}. Therefore, discrepancies observed at earlier times cannot be attributed to finite-size effects.

NLCE and iPEPS results agree up to $t\simeq 0.8$. Given this agreement, together with the internal consistency checks reported for both methods, we use their common result as a reliable benchmark against which to assess the other methods.

We first focus on times $t < t_{\nmax} \simeq 0.4$, where the recursion method is essentially exact. In this regime, its result coincides with that of NLCE/iPEPS. At the same time, small discrepancies are visible in the NG, CTWF and SPD results, at the level of a few times $10^{-3}$, as previously noted in Ref.~\cite{Park_2025_Simulating}. For NG and SPD, these discrepancies appear as systematic shifts, whereas for CTWF they manifest as an apparent noise. In the case of SPD, the deviation is primarily due to the finite time step and can be reduced by decreasing $\Delta t$~\cite{Park_2025_Simulating}. However, for fixed computational resources, this necessitates increasing $\delta$, which degrades accuracy at later times, as seen in Fig.~\ref{fig comp}. For NG, the discrepancy is likely due to limited variational expressivity of the neural quantum states employed~\cite{Park_2025_Simulating}.

At later times, up to $t \simeq 0.8$, the results of SPD and NG remain  consistent with the NLCE/iPEPS benchmark within the small discrepancies described above, while the CTWF and recursion-method results deviate by a larger amount, on the order of a few times $10^{-2}$.  Importantly, the deviation of the recursion-method result from this consolidated curve remains within its estimated error bounds.

A similar comparison between the recursion method, on the one hand, and NLCE~\cite{Richter_2020_Quantum} and SPD~\cite{Begusic_2025_unpublished}, on the other hand, is shown in Fig.~\ref{fig 2D}(b2) for $h_z=1$. The same pattern is observed: good agreement for $t \lesssim t_{\nmax}$, followed by modest deviations at later times of order a few times $10^{-2}$, again well within the estimated error window. Remarkably, for this value of $h_z$, the numerically exact results for the finite $5\times5$ system are already close to the thermodynamic-limit result up to quite large times $t\lesssim 1$.

Finally, we turn to the three-dimensional case. Here, SPD provides \cite{Begusic_2025_Real-time}, to the best of our knowledge, the only independent benchmark currently available. The comparison, shown in Fig.~\ref{fig 3D}(b2), is qualitatively similar to the two-dimensional case but exhibits smaller discrepancies. This suggests that the error estimate in three dimensions may be conservative, and that the recursion method may remain accurate beyond $t_{\nmax}$ to a greater extent than the worst-case bound indicates.

\section{Quench in a one-dimensional system \label{sec: 1D}}

\begin{figure}[t!] 
		\centering
    	\includegraphics[width=\textwidth]{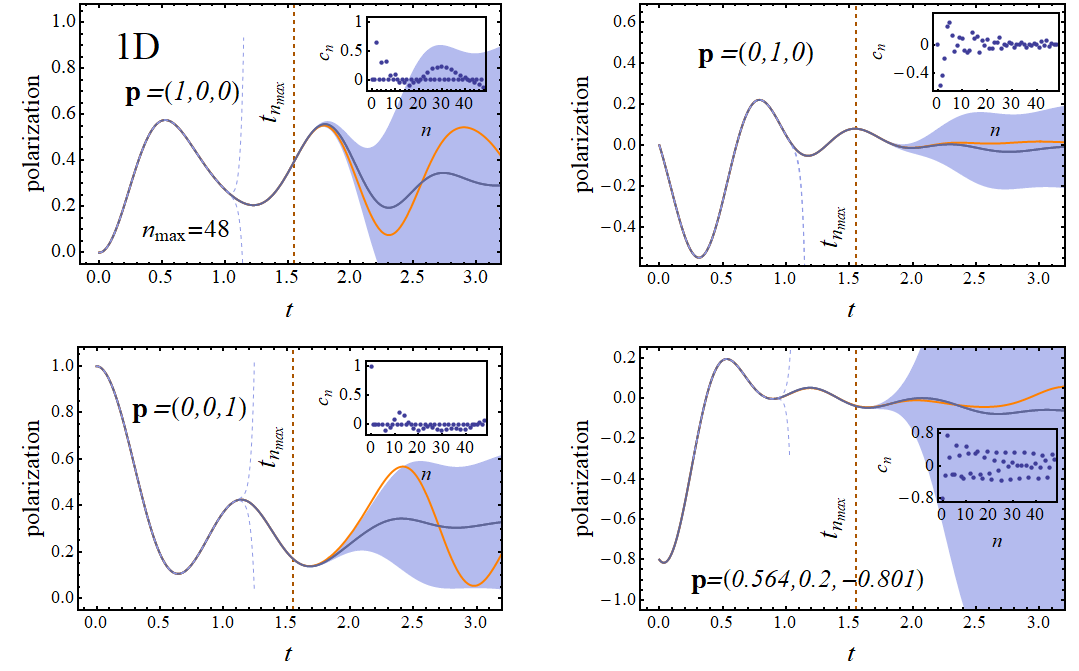}
		\caption{
Time evolution of the total polarization along the $z$ direction in the one-dimensional Ising model~\eqref{H 1D Ising} with $h_x = h_z = 1$. The initial state is a translation-invariant product state~\eqref{initial state} with polarization $\p = (p_x, p_y, p_z)$ indicated in each panel. As in previous figures, the blue line denotes the recursion-method result, with the shaded area indicating the estimated uncertainty, and dashed blue lines show the truncated Taylor expansion for truncation orders $\nmax$ and $(\nmax - 2)$. For $\p = (0,1,0)$, the two dashed lines are visually indistinguishable due to an accidental near-coincidence.  The orange line shows essentially exact results of numerical time evolution (see text for details). Insets display the quench coefficients $c_n$.
}
		\label{fig 1D}
\end{figure}

In one dimension, quench dynamics can typically be computed with high accuracy over relatively long times using matrix-product-state-based methods \cite{Schollwock_2011_Density-matrix,Fishman_2022_ITensor,Fishman_2022_Codebase}. The purpose of applying the recursion method in one dimension is therefore not to generate new data, but rather to benchmark it against essentially exact results. This is the goal of the present section.

We note that the recursion method has recently been applied to one-dimensional quench dynamics in Ref.~\cite{Loizeau_2025_Opening}.

We consider the paradigmatic tilted-field Ising  spin-$1/2$ chain with the Hamiltonian
\begin{equation}\label{H 1D Ising}
H=-\sum_j( \sigma^x_{j}\sigma^x_{j+1}+h_z  \sigma^z_{j}+ h_x \sigma^x_{j}).
\end{equation}
This model is a close kin of the higher-dimensional transverse-field Ising model \eqref{H Ising}; however, it includes an additional field component $h_x$ in order to break integrability otherwise present in the one-dimensional transverse-field Ising model. The observable is the same as above: the total polarization in the $z$-direction, defined in eq.~\eqref{observable}. We reached $\nmax=48$ for this model.

In one dimension, the leading scaling of Lanczos coefficients acquires a logarithmic correction \cite{Parker_2019}. Taking this into account, we use the following extrapolation formula ({\it cf.} eq. \eqref{b_n extrapolation 2D}):
\begin{align}\label{b_n extrapolation 1D}
  b_n &\simeq \frac{\alpha\, n}{\log n} + \gamma + \frac{(-1)^n \tilde\alpha }{\log n+c}.
\end{align}
The rest of the recursion method is implemented as in higher dimensions. The results for four different initial states are shown in Fig. \ref{fig 1D}.

We benchmark the recursion-method results against essentially exact numerics obtained via exact diagonalization using QuSpin \cite{quspin} for system sizes $14$ to $16$, and time-dependent variational principle with matrix product states implemented in ITensor \cite{Fishman_2022_ITensor,Fishman_2022_Codebase} for system size $40$. We have verified that, on the timescale shown in Fig.~\ref{fig 1D}, these results are converged to the thermodynamic limit.

The comparison confirms a major conclusion of Sec.~\ref{sec: quench}: The recursion method is essentially exact up to $t_{\nmax}$, but may deviate from the true value at later times. Fig.~\ref{fig 1D} also confirms both that the error estimate \eqref{truncation uncertainty} indeed provides an upper bound on the actual discrepancy and that this bound may be overly conservative.

We further observe that, as in higher dimensions, the quench coefficients in one dimension display a wide variety of behaviors and no clear universal pattern, see Fig.~\ref{fig 1D}. In contrast to higher dimensions, however, we do not encounter sequences whose magnitude grows beyond unity.

Finally, we note that, unlike in two and especially three dimensions, the time scale accessible to the recursion method in one dimension remains significantly shorter than the thermalization time scale, even for favorable initial states with rapidly decaying quench coefficients. This is likely a consequence of the slower thermalization typical for one-dimensional systems. In particular, the tilted-field Ising model \eqref{H 1D Ising} is known to exhibit anomalous thermalization behavior \cite{Banuls_2011_Strong,Lin_2017_Quasiparticle,Kormos_2017_Real-time,Peng_2022_Bridging,Pathak_2025_Relaxation,Jiang_2026_Krylov} (see also \cite{Bayocboc_2025_L100406} for related results). Moreover, initialization in a translation-invariant product state is known to further slow down thermalization \cite{Ermakov_2022_Thesis,Ermakov_2025_Classical,Pizzi_2025_Genuine}.

In the previous version of this manuscript \cite{previous_version}, we linked the breakdown of the recursion method to a dynamical quantum phase transition (DQPT), motivated by an apparent coincidence between the DQPT time and the onset of deviations of the recursion-method result from the true value. A more careful analysis shows that this is likely a mere  coincidence which, furthermore, does not hold across most of the model parameter space. We therefore retract our earlier claim that a DQPT is the cause of, or a universal precursor to, the breakdown of the recursion method in one dimension. Additional data and discussion on this point is provided in the Appendix \ref{appendix: DQPT}.



\section{Summary and outlook \label{sec: summary}}

To summarize, we have extended the recursion method to quantum many-body quench dynamics and examined its strengths and limitations in this far-from-equilibrium setting. We have performed case studies of two- and three-dimensional transverse-field Ising models and benchmarked the method against essentially exact results in the one-dimensional tilted-field Ising model.

We have established that the method yields essentially exact results up to a time $t_{\nmax}$, which scales logarithmically with the maximal number $\nmax$ of explicitly computed nested commutators and Lanczos coefficients. Beyond this time, the results begin to deviate from the true values, either gradually or abruptly, depending on the initial state.

We have shown that the main limitation of the recursion method for quench dynamics arises from the nonuniversal behavior of the quench coefficients $c_n$ -- the coefficients of the expansion of the initial state $\rho_0$ in the operator Lanczos basis. In contrast to the Lanczos coefficients $b_n$, which exhibit universal behavior and can be reliably extrapolated, the quench coefficients display no universal structure as a function of $n$ and, in general, cannot be extrapolated beyond $\nmax$. This makes the computation of quench dynamics significantly more challenging than that of correlation functions, the latter depending only on $b_n$.

We have demonstrated a broad range of behaviors for the sequences $c_n$, from decay to irregular behavior and even unbounded growth. We find that for initial states corresponding to decaying sequences, the recursion method can remain accurate well beyond $t_{\nmax}$, in some cases approaching the thermalization timescale. In contrast, for initial states with  growing quench coefficients, the method typically loses accuracy immediately beyond $t_{\nmax}$.

We note that the irregular and potentially divergent behavior of the quench coefficients may limit the applicability of other concepts and techniques based on the recursion method \cite{Parker_2019,Chu_2024_Quantum,Loizeau_2025_Opening,Alishahiha_2025_Thermalization,rabinovici2025krylov}, and should therefore be carefully taken into account.

Finally, we have compared the practical performance of the recursion method in two and three dimensions with other state-of-the-art approaches to quench dynamics. In two dimensions, the conclusions are as follows. For times $t < t_{\nmax}$, the recursion method yields highly accurate results that coincide with the benchmark established by the numerical linked-cluster expansion (NLCE)~\cite{Richter_2020_Quantum} and the tensor-network approach (iPEPS)~\cite{Dziarmaga_2022_Time}. The other methods considered -- neural-network approaches~\cite{Sinibaldi_2026_Time-Dependent,Chen_2026_Convolutional} and sparse Pauli dynamics (SPD)~\cite{Begusic_2025_Real-time} -- yield results that remain close to the NLCE/iPEPS benchmark, but exhibit systematic discrepancies on the order of $10^{-3}$. At the same time, NLCE~\cite{Richter_2020_Quantum}, iPEPS~\cite{Dziarmaga_2022_Time}, SPD~\cite{Begusic_2025_Real-time}, and NG~\cite{Sinibaldi_2026_Time-Dependent} extend to substantially longer times than $t_{\nmax}$ while remaining mutually consistent within these small discrepancies, whereas the recursion method shows noticeable deviations for $t \gtrsim t_{\nmax}$.

We emphasize, however, that among these methods the recursion method is unique in producing symbolic results that cover the entire model parameter space and arbitrary explicit initial states in a single computation. This feature is particularly valuable given the substantial computational cost of these simulations. As a result, the recursion method stands out in a distinct niche: it is especially useful when accurate results at moderate times are required across many parameter values and/or initial states.

In three dimensions, SPD~\cite{Begusic_2025_Real-time} provides the only available benchmark. For initial states with favorable behavior of quench coefficients, we find that the recursion-method results essentially coincide with those of SPD for $t < t_{\nmax}$ and remain reasonably consistent at later times, with discrepancies on the order of a few times $10^{-2}$. Notably, in this regime the recursion method can reach times close to the thermalization timescale. We therefore conclude that the recursion method is particularly promising in three dimensions.

We conclude by outlining two directions for further development of the recursion method for quench dynamics. The first is to extend the computation of quench coefficients beyond $\nmax$ for specific classes of initial states by exploiting their structure. For instance, for an uncorrelated state polarized along the $z$ direction, all Pauli strings in the Heisenberg operator that contain $x$- or $y$-Pauli matrices do not contribute to the expectation value. This observation allows one to discard Pauli strings with a large number of $x$- and $y$-Pauli matrices during the computation of nested commutators, thereby enabling the evaluation of more quench coefficients at the same computational cost. This idea has already been implemented in the sparse Pauli dynamics approach~\cite{Begusic_2025_Real-time}, where it proved instrumental in extending simulations to much longer times.

A second promising direction is to combine the recursion method with approaches formulated in the Schr\"odinger picture. In such a hybrid scheme, the state of the system is first evolved using a Schr\"odinger-picture method up to the maximal accessible time, and then the result is used as the initial condition for the recursion method, which extends the evolution further~\cite{rudolph2025pauli,xu2026classical}. Notably, the Schr\"odinger part of this hybrid evolution can be carried out on quantum hardware~\cite{Fuller_2026_Improved}.


\section*{Acknowledgements}

We are grateful to Tomislav Begu\v{s}i\'{c} and Garnet Kin-Lic Chan for discussing their paper \cite{Begusic_2025_Real-time} and our manuscript, and in particular for providing unpublished data \cite{Begusic_2025_unpublished} on the 2D Ising model. We thank Berislav Bu\ifmmode \check{c}\else \v{c}\fi{}a and Nicolas Loizeau for useful remarks and discussions, and Luka Pave\v{s}i'{c} for discussing ref.~\cite{Pavesic_2025_Constrained}. We are grateful to Jacek Dziarmaga, Alessandro Sinibaldi and Robin Steinigeweg for sharing data from refs.~\cite{Dziarmaga_2022_Time}, \cite{Sinibaldi_2026_Time-Dependent} and \cite{Richter_2020_Quantum}, respectively, and to Markus Heyl for sharing and discussing data from ref.~\cite{Chen_2026_Convolutional}. Finally, we gratefully acknowledge numerous discussions with Alexander Teretenkov and Nikolay Il'in.

\paragraph{Author contributions} OL initiated the research, analyzed the data, and wrote the manuscript. ISh, VKh, FU, and IE computed the Lanczos and quench coefficients. ID carried out the benchmark calculations for the one-dimensional model.

\paragraph{Funding information}
This work was supported by the Russian Science Foundation under grant N~24-22-00331 \url{https://rscf.ru/en/project/24-22-00331/}

\appendix

\section{Zero energy shell \label{appendix: zero energy}}

\begin{figure}[t] 
		\centering
    	\includegraphics[width=\linewidth]{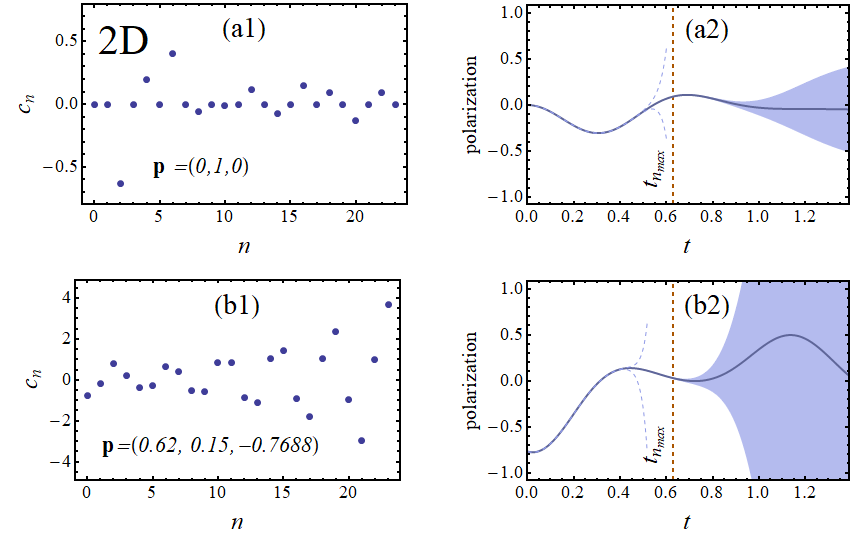}
		\caption{The same as in Fig. \ref{fig 2D}(b1)--(c2), but for initial states from the energy shell with zero energy, corresponding to infinite temperature. }
		\label{fig 2D E0}
\end{figure}

\begin{figure}[t] 
		\centering
    	\includegraphics[width=\linewidth]{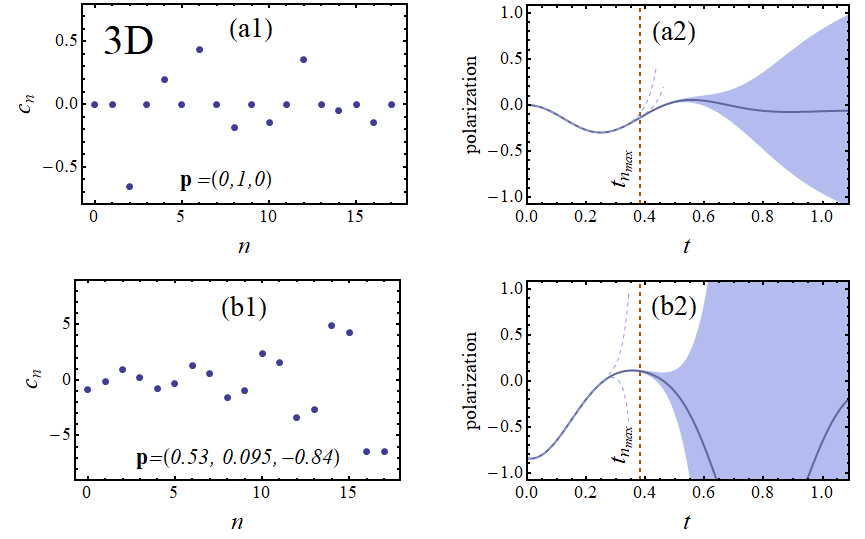}
		\caption{The same as in Fig. \ref{fig 3D}(b1)--(c2), but for initial states from the energy shell with zero energy, corresponding to infinite temperature. }
\label{fig 3D E0}
\end{figure}

Here we provide additional results for initial states with zero average energy, see Figs. \ref{fig 2D E0},\ref{fig 3D E0}. This energy shell corresponds to the infinite temperature. One can see that the results are qualitatively similar to those discussed in the main text and shown in Fig.~\ref{fig 2D},\ref{fig 3D}.

\begin{figure}[t] 
		\centering
    	\includegraphics[width=\linewidth]{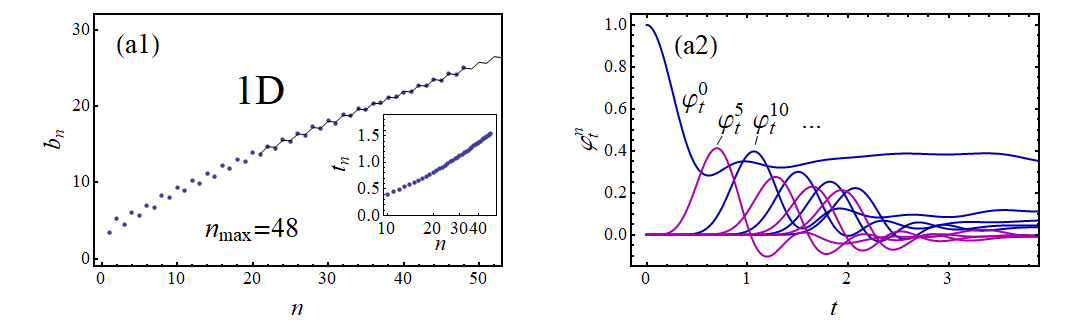}
		\caption{The same as in Fig. \ref{fig 2D}(a1),(a2), but for the one-dimensional tilted-field Ising model \eqref{H 1D Ising} with $h_x=h_z=1$. }
\label{fig 1D Lanczos coefficients}
\end{figure}

\section{Moments and Lanczos coefficients}

In Tables \ref{table 2D Ising}, \ref{table 3D Ising} and \ref{table 1D Ising} we provide values of moments $\mu_{2n}=(A|\L^{2n} A)/ (A|A)$ and Lanczos coefficients $b_n$ for the transverse-field Ising model \eqref{H Ising} in, respectively, two and three dimensions, and for the one-dimensional tilted-field Ising model \eqref{H 1D Ising}. An algorithm to obtain the Lanczos coefficients from the moments can be found in \cite{Dupuis_1967_Moment,viswanath2008recursion}.  Additionally, we plot the Lanczos coefficients and function $\varphi^n_t$ for the one-dimensional case in Fig.~\ref{fig 1D Lanczos coefficients}.

Note that in the previous version of the manuscript \cite{previous_version}, for the 3D Ising model, we provided 12 moments and Lanczos coefficients for the non-translation-invariant observable $\sigma^z_j$. Here we provide 17 moments and Lanczos coefficients for the respective translation-invariant observable. While the values are different in the two cases, the output of the recursion method is consistent whenever the initial state is translation-invariant.

\begin{table}[h]

\begin{tabular}{l|l|l}
    {$n$} & {$\mu_{2n}$} & $b_n$ \\
    \hline
    1  & 12 & 3.4641 \\
    2  & 480 & 5.2915 \\
    3  & 25984 & 4.4934 \\
    4  & 1694208 & 6.0281 \\
    5  & 127258624 & 5.6190 \\
    6  & 10783342592 & 6.9438 \\
    7  & 1019673509888 & 6.6972 \\
    8  & 107060323680256 & 8.2286 \\
    9  & 12469034121428992 & 7.9953 \\
    10 & 1613762515315982336 & 9.2837 \\
    11 & 232736621970029805568 & 8.9148 \\
    12 & 37468646645944075419648 & 10.2773 \\
    13 & 6726394693098760675786752 & 9.8677 \\
    14 & 1340433610285320688734568448 & 11.1094 \\
    15 & 294391435294882867890188451840 & 10.7766 \\
    16 & 70659698635880317667608199430144 & 12.1889 \\
    17 & 18383424921170050832257258778263552 & 11.7912 \\
    18 & 5147501922089241875609891362536685568 & 12.9532 \\
    19 & 1542402400283276113669614948668265201664 & 12.7320 \\
    20 & 492432525422001580271341582388177351475200 & 13.9288 \\
    21 & 166985805198508106978965686562725967100379136 & 13.5778 \\
    22 & 60011852488095956067395190665735924993040056320 & 14.6867 \\
    23 & 22821321744324019152247154618657381361730948431872 & 14.4673 \\
    24 & 9172207374041781749043802673339425314176975748726784 & 15.5798 \\
    25 & 3892052853514868229060585022861205953766446860739805184 & 15.3892 \\
    26 & 1741736488569588529172771998235278437950082051065922977792 & 16.4064 \\
    27 & 821019133269536743838893539994338673397100678325485105577984 & 16.1096 \\
    28 & 407086843763398331397385204208407255958886525287040747053252608 & 17.3132 \\
    29 & 211990829889808789195437699387039831255135134538810758197371994112 & 17.0171 \\
    30 & 115754651941408887232253140259122456149371760683740727687331373383680 & 18.0978 \\
    31 & 66166596237703872256126915723522220984465097504279554678985792092635136 & 17.7211 \\
    32 & 39530072936257449901374525090449349559603851447100658065034700516313006080 & 18.8846 \\
\end{tabular}

\caption{Moments and Lanczos coefficients for the 1D Ising model \eqref{H 1D Ising} at $h_x=h_z=1$ and observable \eqref{observable}. Moments for $n=33,\dots,48$ are too large to fit the page. They can be obtained from the authors upon a request. \label{table 1D Ising}}
\end{table}

\begin{table}[h]
\begin{tabular}{l|l|l}
    {$n$} & {$\mu_{2n}$} & $b_n$ \\
    \hline
    1  & 16 & 4.0000 \\
    2  & 896 & 6.3246 \\
    3  & 71680 & 5.7966 \\
    4  & 7307264 & 7.9714 \\
    5  & 908099584 & 7.7771 \\
    6  & 134882656256 & 9.8932 \\
    7  & 23708042264576 & 9.8014 \\
    8  & 4897950853496832 & 11.7753 \\
    9  & 1181552437773729792 & 11.8055 \\
    10 & 330379916698579894272 & 13.6724 \\
    11 & 106222899499818563928064 & 13.7546 \\
    12 & 38952574279318656784531456 & 15.6436 \\
    13 & 16166117065306264371967557632 & 15.7188 \\
    14 & 7539715274735761761366362292224 & 17.3406 \\
    15 & 3926354218204565053083857654382592 & 17.3814 \\
    16 & 2269213981250751881428716107876794368 & 19.0685 \\
    17 & 1446793491966272434190827088127549505536 & 19.2419 \\
    18 & 1011459407804179014361425714300323771711488 & 20.8675 \\
    19 & 770720014993499216420856132351984701670424576 & 21.0498 \\
    20 & 636516730341979806791563685949386538476858507264 & 22.6354 \\
    21 & 566952387953410930810990384866974059612775307542528 & 22.8705 \\
    22 & 542434116478048005253046423594975291775134675528318976 & 24.4221 \\
    23 & 555699452970039450718350263160732485341995086705577689088 & 24.6588 \\
\end{tabular}
\caption{Moments and Lanczos coefficients for the observable~\eqref{observable} and the Ising model on the square lattice, eq. \eqref{H Ising}, with $h_z=1$.\label{table 2D Ising}}
\end{table}

\begin{table}[h]
\begin{tabular}{l|l|l}
    {$n$} & {$\mu_{2n}$} & $b_n$ \\
    \hline
    1  & 24 & 4.89898 \\
    2  & 1920 & 7.48331 \\
    3  & 230400 & 7.55929 \\
    4  & 36618240 & 9.98284 \\
    5  & 7313326080 & 10.1916 \\
    6  & 1790621122560 & 12.7955 \\
    7  & 530848042450944 & 13.1431 \\
    8  & 188884569821282304 & 15.4032 \\
    9  & 79892590021627084800 & 15.8832 \\
    10 & 39721461646806598287360 & 18.2052 \\
    11 & 22947716425271653039079424 & 18.7155 \\
    12 & 15244534918363394241158184960 & 20.9982 \\
    13 & 11543604835131049187075722051584 & 21.5619 \\
    14 & 9891579029630566720438258058133504 & 23.8488 \\
    15 & 9532138756054379075322538513856987136 & 24.4597 \\
    16 & 10273436716844388765436471742460667625472 & 26.751 \\
    17 & 12321137907981191488465349016456502941057024 & 27.4636 \\
\end{tabular}
\caption{Moments and Lanczos coefficients for the observable~\eqref{observable} and the Ising model on the cubic lattice, eq. \eqref{H Ising}, with $h_z=1$.\label{table 3D Ising}}
\end{table}

\section{Recursion method breakdown and dynamical quantum phase transition in one dimension \label{appendix: DQPT}}

\begin{figure}[t] 
		\centering
    	\includegraphics[width=\linewidth]{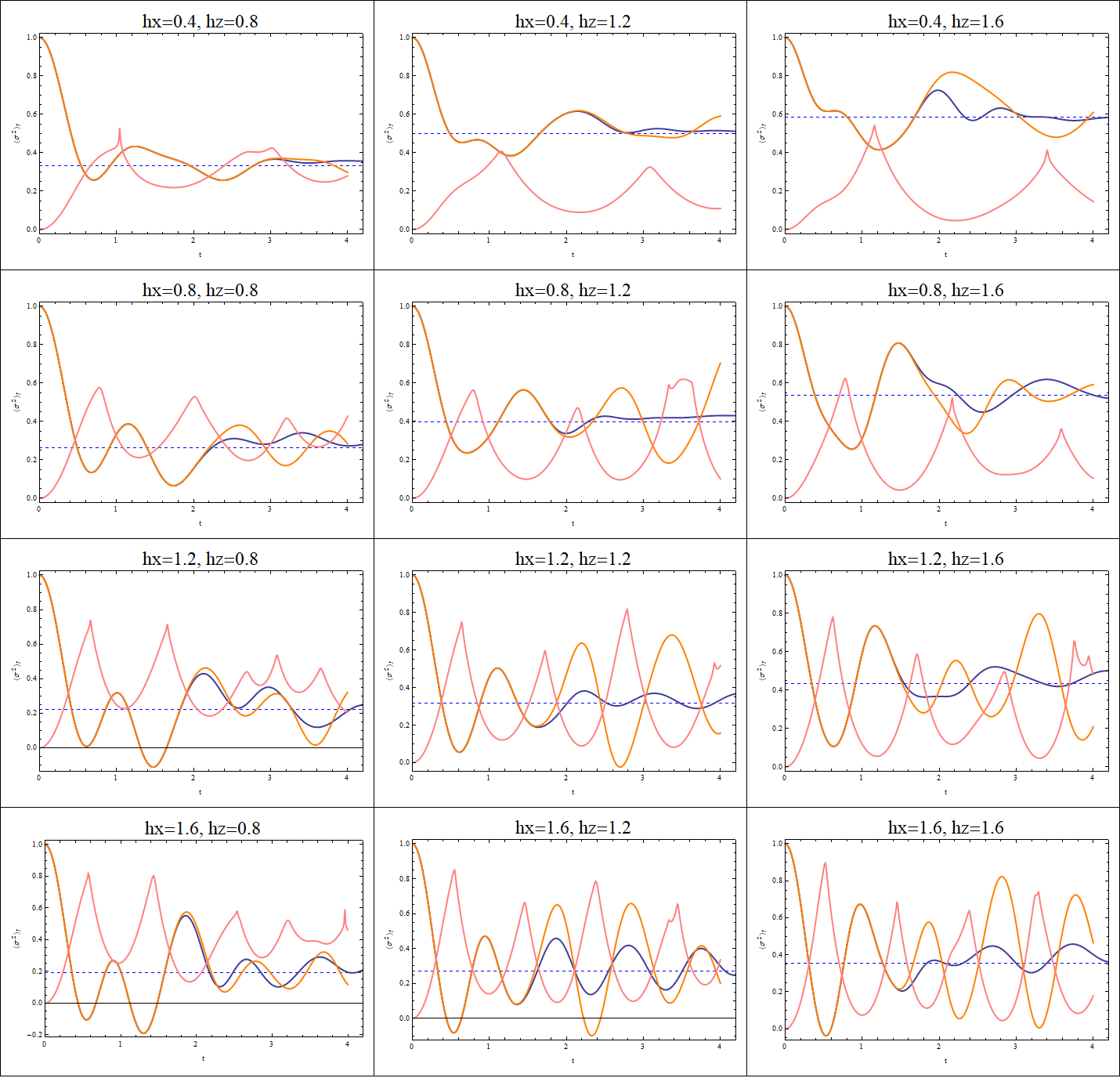}
		\caption{Time evolution of the total polarization along $z$ direction in the 1D Ising model \eqref{H 1D Ising} for various values of $h_z$ and $h_x$. The initial state is a product translation-invariant state \eqref{initial state} polarized in the $z$-direction.  Orange line -- essentially exact results of the numerical time evolution (see text for details). Blue line -- results of the recursion method. Pink line -- the rate function~$\lambda(t)$ whose cusps mark the dynamical quantum phase transitions.  }
		\label{fig_1D_Ising_supp}
\end{figure}

In the previous version of this manuscript \cite{previous_version}, we  attributed the breakdown of the recursion method to a dynamical quantum phase transition (DQPT) \cite{Pollmann_2010_DQPT,Heyl_2013_Dynamical,Karrasch_2013_Dynamical,Heyl_2018_Dynamical} -- a phenomenon by which the Loschmidt rate function $\lambda(t)=-N^{-1}\,\log\big|\langle \Psi_0|e^{-i\,t\,H}|\Psi_0\rangle\big|^2$ features singularities in the thermodynamic limit.\footnote{
There is an ambiguity in the meaning of the term ``dynamical (quantum) phase transition.'' Historically, it was introduced in a quite different context, referring to a transition between distinct phases of matter that unfolds over time due to variations in the Hamiltonian \cite{Schutzhold_2006_Sweeping,Eckstein_2009_Thermalization,Uhlmann_2010_System,Garrahan_2010_Thermodynamics,Diehl_2010_Dynamical,Sciolla_2010_Quantum} -- a definition more closely aligned with the conventional understanding of phase transitions. In contrast, the definition we adopt was introduced later \cite{Heyl_2013_Dynamical} and is based on a formal analogy between the Loschmidt rate and the free energy.}

This conclusion was motivated by an apparent coincidence between the DQPT time and the onset of deviations of the recursion-method result from the true value. A more careful analysis presented in Fig.~\ref{fig_1D_Ising_supp}  shows that this is likely a mere  coincidence which, furthermore, does not hold across most of the model parameter space. The apparent agreement occurs, within uncertainties, only for $h_x \simeq h_z$. We therefore retract our earlier strong claim regarding a relation between the recursion method breakdown and the DQPT. For a discussion of DQPT in the context of the Lanczos method in the Schr\"odinger representation, see Ref.~\cite{Takahashi_2025_Dynamical}.

\bibliography{C:/D/Work/QM/Bibs/recursion_method,C:/D/Work/QM/Bibs/DQPT,C:/D/Work/QM/Bibs/other_methods,C:/D/Work/QM/Bibs/thermalization,C:/D/Work/QM/Bibs/scars}


\nolinenumbers

\end{document}